\documentclass[reprint, amsmath,amssymb, aps]{revtex4-2}

\usepackage{graphicx}
\usepackage{comment}
\usepackage{color}
\usepackage{hyperref}
\usepackage[utf8]{inputenc}

\begin{document}

\preprint{APS/123-QED}

\title{Real-time detection of Rydberg state dynamics of cold atoms using an optical cavity}

\author{Elmer Suarez}
\author{Philip Wolf}
\author{Patrizia Weiss}
\author{Sebastian Slama}
\email{sebastian.slama@uni-tuebingen.de}
\affiliation{Center for Quantum Science and Physikalisches Institut, Eberhard-Karls Universität Tübingen, Auf der Morgenstelle 14, 72076 Tübingen, Germany }

\date{\today}

\begin{abstract} 
This work reports on the real-time detection of internal-state dynamics of cold Rb$^{87}$ atoms being excited to the $30D_{5/2}$ Rydberg state via two-photon excitation. A mesoscopic cloud of atoms is overlapped with the mode volume of a confocal optical cavity and optically pumped by two laser beams transverse to the cavity axis. The excitation to Rydberg states changes the collective atom-cavity coupling, which is detected by monitoring the light transmitted through the cavity while being weakly driven. In addition to the damped coherent excitation dynamics and the decay back to the ground state, the data show a superradiant enhancement of the black-body radiation induced transitions from the $30D_{5/2}$ state to neighboring Rydberg states. Furthermore, they show a density dependent mitigation of the superradiant decay which is attributed to long range dipole-dipole interactions between atoms in the involved Rydberg states. These results contribute to solving a recent controversy on the interplay between BBR-induced superradiance and Rydberg atom interactions. 
\end{abstract}

\maketitle


\section{\label{sec:intro}Introduction}
Rydberg atoms are hot candidates for realizing quantum gates \cite{Wilk2010,Isenhower2010}, quantum memories \cite{Li2016}, single-photon devices \cite{Gorniaczyk2014,Tiarks2014}, and long-range interacting many-body quantum systems \cite{Schauss2015}. A key step for quantum experiments with Rydberg atoms is the coherent excitation of Rydberg states with high excitation probability \cite{Cubel2005, Deiglmayr2006}. In this context, atomic ensembles in optical cavities are of particular interest due to the collective enhancement of light-matter interaction. Since the collective single photon Rabi frequency scales as $g_\mathrm{coll}=g_0\sqrt{N}$ with coupling constant $g_0$ and atom number $N$, the required time for realizing a $\pi$-pulse decreases with $1/\sqrt{N}$ and thus makes it easier to overcome decoherence and loss.\\
Ensembles, however, suffer from shot-to-shot atom number fluctuations that limit the efficiency of the excitation. For that reason, excitation with high probability requires the precise knowledge of the atom number, or real-time monitoring of the excitation dynamics. In this regard, the cavity transmission can serve as a monitor for the number of atoms. For a sufficiently high atom number resolution, this can even be used for spin squeezing \cite{Leroux2010,Chen2011}. Recently, the transmission of an optical cavity has been used to monitor the expansion of an atomic cloud \cite{Niederriter2020}, a dynamical phase transition \cite{Clark2021} and evaporative cooling \cite{zeiher21}, all in real-time. These approaches are based on the frequency shift caused by the atoms in the dispersive limit when the cavity is far detuned from atomic resonance. A complementary approach detecting the transmission through a superconducting microwave cavity to count Rydberg atoms has also been demonstrated recently \cite{Garcia2019}. Note, that real-time detection of Rydberg excitation has also been observed in hot vapor cells by detecting the transmission of the laser driving the lower transition of the two-photon process \cite{Huber2011,Baluktsian2013}.\\
The coherent excitation of Rydberg atoms is additionally complicated by the fact that Rydberg states are coupled by room-temperature black-body radiation (BBR) to adjacent Rydberg levels, spreading the occupation over several states \cite{archimi2019}. Furthermore, superradiance (SR) can lead to a speed up of this spreading, as observed with atomic beams in a superconducting microwave cavity \cite{raimond1982}, or triggered in free space by a weak microwave pulse  \cite{grimes2017}. However, BBR-induced superradiance in clouds of cold atoms has been controversially discussed in the Rydberg community, as some experiments have reported its direct or indirect observation \cite{wang2007, Weatherill_2008, JODay2008, Hao2021}, whereas others have reported its absence \cite{Zhou2016}. A recent theoretical work \cite{sutherland2017} has shed light on this dispute by considering the influence of dephasing induced by dipole-dipole interactions on superradiance, and the competition between superradiant transitions.\\
The present work reports on the real-time detection of Rydberg state dynamics of cold Rb$^{87}$ atoms including the excitation to the Rydberg state and the decay back to the ground state. In particular, we experimentally investigate the influence of black body radiation, spreading the population to adjacent Rydberg states, and the role of superradiance and dipole-dipole interactions in this process. The real-time detection is facilitated by observing the transmission through an optical cavity tuned in resonance with the D2 line transition. Here, the atom-cavity interaction is governed by stimulated absorption and emission of cavity photons. This makes the cavity blind to atoms in states not interacting with cavity photons: Rydberg states for this work. Therefore, detecting the cavity transmission is a real-time monitor for Rydberg dynamics: from the coherent excitation to its redistribution to neighboring states and, therefore, to Rydberg-Rydberg dynamics.

\section{\label{sec:theory}Theoretical background}
It is convenient to separate the total system in two parts, each considering different types of atom-light interactions. In Sec. \ref{CavityDyn} the atom-cavity interaction is introduced and simplifications are justified. Sec. \ref{RydDyn} introduces the atomic dynamics arising from the interaction with both transverse classical fields. At the end we show that it is possible to use the adiabatic approximation to unite both parts.
\subsection{Atom-cavity system}\label{CavityDyn}
\begin{figure}[t]
	\includegraphics{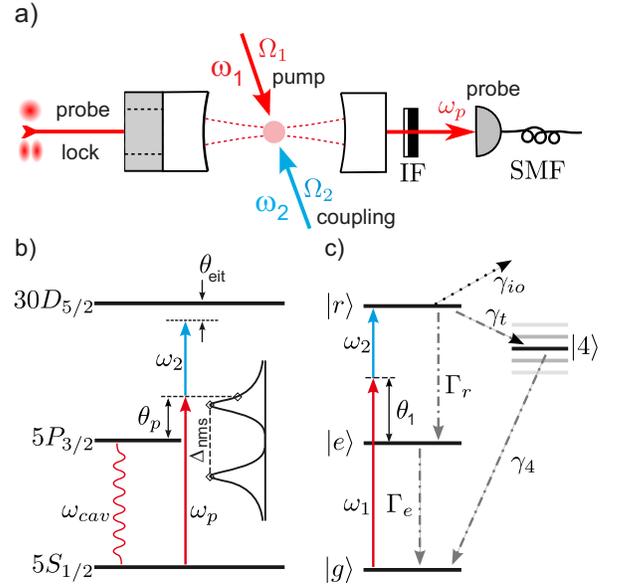}
	\caption{\label{fig1} Experimental scheme: a) Two transverse lasers with frequencies $\omega_1$ and $\omega_2$ excite atoms in the cavity to Rydberg states. A longitudinal laser with frequency $\omega_p$ probes simultaneously the atom-cavity system. The transmitted probe light transmitted through the cavity is filtered (IF=interference filter) from the lock light used to stabilize the cavity length, coupled into a singe mode fiber (SMF), and detected by an avalanche photodiode. b) Relevant atomic levels of the atom-cavity interaction. The cavity resonance frequency is tuned in resonance with the lower transition generating a symmetric normal mode splitting $\Delta_\mathrm{nms}=2g_0\sqrt{N}$. The probe laser frequency is tuned to the slope of one of the normal modes. c) Atomic levels and decay rates involved in the Rydberg excitation.}
\end{figure}

The description of our system starts considering a collection of two-level atoms interacting with a single cavity mode being driven by an external field, Fig.~\ref{fig1}a) and b). In this context, the single-atom cooperativity is introduced:
\begin{equation}\label{cooperativity}
C_1=\frac{2g_0^2}{\kappa\Gamma_e},   
\end{equation}
Where $\Gamma_e$ and $\kappa$ are the decay rates of the excited state and the cavity field, respectively. As will be explained in more detail in the next section, several thousands of atoms are interacting with the cavity and, at the same time, $C_1$ is on the order of $10^{-3}$. Therefore, we apply the semi-classical approximation  neglecting atom-photon correlations when writing the mean-field equations. This leads us to a widely used set of Maxwell-Bloch equations \cite{Orozco1997}.\par
In order to calculate the intra-cavity photon number, we  consider the driving field with frequency $\omega_p$ to be weak and introduce a probability $P=P_e+P_g$ for an atom to be in one of the two states $\left|g\right>$ or $\left|e\right>$, i.e. the states that are coupled by the cavity field. The probability for the atom to be in any other state non interacting with the cavity field, in particular a Rydberg state, is correspondingly $1-P$. Then, the steady-state for the normalized cavity photon number is given by
\begin{equation}\label{CavTrans}
    V=\left\vert \frac{1}{1+i\left( \Delta-2C_1N_\mathrm{eff}P\chi\right)} \right\vert ^2,
\end{equation}
where $\Delta=2\left(\theta_\mathrm{cav}-\theta_p\right)/\kappa$, with $\theta_\mathrm{cav}$ and $\theta_p$ the cavity and driving field detuning, relative to atomic resonance, $N_\mathrm{eff}$ the effective atom number inside the cavity mode volume, and $\chi=i/\left(1-i2\theta_p/\Gamma_e\right)$ a factor proportional to the atomic susceptibility.\par
\begin{figure}[t]
\includegraphics{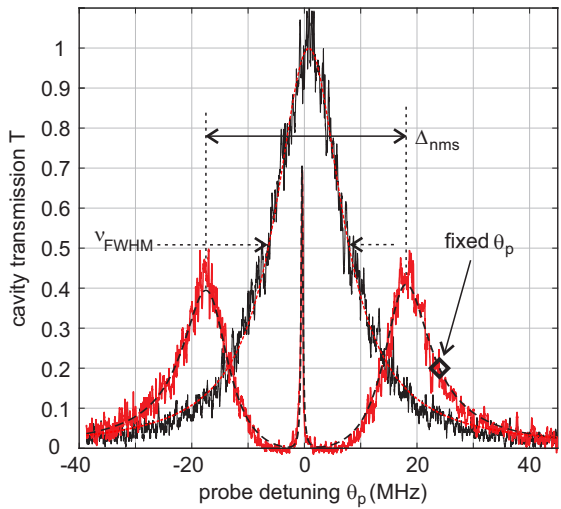}
\caption{\label{fig2} Measured cavity transmission spectrum of the empty cavity (black data) and its fit (red dashed line) with full width at half maximum of $\nu_\mathrm{FWHM}=13.4~\mathrm{MHz}$, and of the cavity filled with atoms and the coupling beam on (red data), featuring a normal mode splitting $\Delta_\mathrm{nms}$, corresponding to 7.300 atoms, and a cavity Rydberg EIT signal. The Rabi frequency of the coupling laser $\Omega_2=2\pi\times5.9~\mathrm{MHz}$ and the broadened linewidth $\gamma_R=2\pi\times100~\mathrm{kHz}$ of the Rydberg state are determined from the theoretical fit (black dashed line). The fixed probe light detuning $\theta_p=2\pi\times24~\mathrm{MHz}$ used for detecting the internal state dynamics is tagged. In each curve, the probe frequency has been scanned within 1.5 ms across the atomic resonance. The curves are normalized to the fitted empty cavity peak transmission.}
\end{figure}

The classical field with Rabi frequency $\Omega_2$ coupling the excited state to the Rydberg state (i.e. the coupling beam) is included into Eq. $\left(\ref{CavTrans}\right)$ by a modified susceptibility:
\begin{equation}\label{Susc3}
    \chi'=\frac{i}{1-i2\theta_p/\Gamma_e+\frac{\Omega_2^2/\gamma_R\Gamma_e}{1-i2\theta_\mathrm{eit}/\gamma_R}}.
\end{equation}
Here, we have introduced the rate $\gamma_R=\Gamma_r+\gamma_d$ with the inverse lifetime $\Gamma_r$ of the Rydberg state, and dephasing rate $\gamma_d$ which will be introduced in the next section, and $\theta_\mathrm{eit}=\theta_p+\theta_2$, with detuning $\theta_2$ of the coupling beam relative to the upper atomic transition. Thus, the case $\theta_{eit}=0$ generates a new resonance, i.e. electromagnetically induced transparency (EIT) through the cavity. Cavity Rydberg EIT has been studied in \cite{Boddeda2016,Sheng2017,Ningyuan2015}. In our experiment, see Fig.~\ref{fig2}, we use it to benchmark the system in terms of $\Omega_2$ and $\gamma_R$. 

\par
\subsection{Two-photon Rydberg excitation}\label{RydDyn}
For a ladder-type system interacting with two classical fields (coupling and pump beam) the Hamiltonian can be written as
\begin{equation}\label{Hamiltonian3}
    H=-\theta_1\vert e\rangle\langle e \vert-\vartheta_\mathrm{tr}\vert r\rangle\langle r\vert+\frac{\Omega_1}{2}(\hat{\sigma}_{ge}+\hat{\sigma}_{eg})+\frac{\Omega_2}{2}(\hat{\sigma}_{er}+\hat{\sigma}_{re}),
\end{equation}
where $\hat{\sigma}_{ij}=\vert i\rangle\langle j\vert$, with $i,j=g,e,r$ (i.e. ground, excited and Rydberg state); $\Omega_1$ is the Rabi frequency for the pump beam, $\vartheta_\mathrm{tr}=\theta_1+\theta_2$ is usually called 2-photon detuning and $\theta_1$ is the detuning of the pump beam relative to atomic resonance, Fig.~\ref{fig1}a) and c). Also, as shown in \cite{Brion_2007}, the intermediate state can be adiabatically eliminated when $\theta_1$ is sufficiently large, and an effective two-level Hamiltonian can be derived. This effective Hamiltonian is parametrized by an effective Rabi frequency 
\begin{equation}\label{OmegaEff}
\Omega_\mathrm{eff}=\frac{\Omega_1\Omega_2}{\theta_1-\theta_2},
\end{equation}
and an effective detuning 
\begin{equation}\label{DeltaEff}
\vartheta_\mathrm{eff}=\vartheta_\mathrm{tr}+\frac{\Omega_1^2-\Omega_2^2}{2(\theta_1-\theta_2)}.
\end{equation}
Stray electric and magnetic fields as well as frequency noise from the lasers contribute to the decoherence of the Rydberg state \cite{Ningyuan2015} with a rate $\gamma_d$ and its action is described by the Lindblad operator
\begin{equation}\label{LindDeph}
    L_d=\sqrt{\gamma_d}\left|r \right>\left< r\right|.
\end{equation}\par
In contrast to the excitation scheme based on the three levels $\left|g\right>$, $\left|e\right>$, and $\left|r \right>$, the decay back to the ground state involves also other states. First of all, even though the main spontaneous decay channel for the Rydberg state is to the $5P_{3/2}$ state, it spontaneously decays also to several other states. Additionally, BBR at room temperature is responsible for transferring the population out to other, neighbouring states. Since neither of the transverse beams interact with these other states, they will only decay to the ground state via spontaneous decay. This problem has been approached in \cite{Zhou2016}, and more simply in \cite{Weatherill_2008,JODay2008}. The latter approach models BBR population transfer out of a Rydberg state by a decay rate $\gamma_t$ to an effective fourth state which will, in turn, decay to the ground state with a rate $\gamma_4$, see Fig.~\ref{fig1}c). In the present work, we take various decay channels into account by a set of rate equations
\begin{equation}\label{RateEqs}
    \frac{d\vec{N}}{dt}=A\vec{N}
\end{equation}
where $\vec{N}$ is a vector containing the populations of all states involved, and $A$ is a matrix containing the transition rates between the states in $\vec{N}$. The Alkali Rydberg Calculator (ARC) \cite{arc} is used to build the matrix $A$ with as many states as wanted. An ionization rate $\gamma_\mathrm{io}$ from the Rydberg state is also included.\par
The $1/n^3$ dependence of the transition frequency between contiguous Rydberg states suggests that these are potential platforms for superradiance. In this context, the transition rate $\Gamma_{rr'}$ between Rydberg states $r$ and $r'$ due to BBR is enhanced by superradiance. To describe this effect, we follow \cite{JODay2008} by adding new terms to the differential equations:
\begin{equation}\label{SupDiff}
    \frac{dN_r}{dt}=-\sum_{l<r} \Gamma_{rl}C_{rl}N_r(N_l+1) + \sum_{l'>r} \Gamma_{rl'}C_{rl'}N_{l'}(N_r+1).
\end{equation}
Here, $r$ represents a Rydberg state while $l$ and $l'$ represent the neighbouring states below and above $r$, respectively. $C_{rl}$ is called the cooperativity parameter and has a value between zero and one, depending on the length scale $R$ of the system and the transition wavenumber $k_{rl}$:
\begin{equation}\label{CoopPar}
    C_{rl}=\frac{9(\sin(k_{rl}R)-k_{rl}R\cos(k_{rl}R))^2}{(k_{rl}R)^6}.
\end{equation}\par
With the Bloch equations governed by the Hamiltonian in Eq. (\ref{Hamiltonian3}) and the Lindblad operator in Eq.~(\ref{LindDeph}), together with Eq.~(\ref{RateEqs}) and (\ref{SupDiff}), we obtain a set of differential equations describing the time dynamics of the 3-level system and all other states in $\vec{N}$. We calculate the cavity field during this evolution by using Eq. (\ref{CavTrans}) under the adiabatic approximation: the atom-cavity system reaches its steady-state on a timescale on the order of 200~ns much faster than the Rydberg dynamics which, for our experimental parameters, is on the order of 2~$\mu s$, and longer. In this way, Rydberg dynamics are introduced into Eq. (\ref{CavTrans}) by the probability $P$ that an atom is either in $\left|g\right>$ or $\left|e\right>$. The resulting time dynamics is then used to analyze our data.

\section{\label{sec:experiment}Experiment}
\subsection{Experimental setup}
Fig.~\ref{fig1}a) shows a schematic of the most relevant parts. A magneto-optic trap (MOT) with a radius of $R_\mathrm{MOT}=0.5~\mathrm{mm}$ ($1/e$-radius of density) is overlapped with the mode of a confocal cavity of 50 mm length (mode beamwaist $w_0=80~\mathrm{\mu m}$). The full width at half maximum $\nu_{FWHM}=13.4~\mathrm{MHz}$ of the empty cavity, see Fig.~\ref{fig2}, corresponds to a finesse $F=224$. The cavity is tuned to be resonant with the $F=2$ to $F'=3$ transition of the D2 line of Rb$^{87}$, i.e. $\theta_\mathrm{cav}=0$. The single atom-cavity coupling equals $g_0=2\pi\times206.8~\mathrm{kHz}$, after taking the average over the Zeeman sublevels. The cavity length is controlled by a piezo-transducer using the Pound-Drever-Hall (PDH) technique with one of the sidebands of a 786 nm ("lock") laser that is amplitude-modulated by an electro-optic modulator (EOM). The carrier frequency of the lock laser is stabilized to a stable high-finesse ultra-low expansion (ULE) cavity serving as a frequency reference. The probe laser frequency is stabilized by Rubidium absorption spectroscopy and shifted by an acousto-optic modulator (AOM).\par
Light from both lock and probe laser are coupled into the cavity using the same beam path, but with orthogonal polarizations. This allows us to discriminate the lock laser in the reflected light when generating the error signal. Furthermore, lock and probe laser are coupled into different transverse modes of the cavity. Light transmitted through the cavity is filtered by an optical bandpass filter and by a single-mode fiber, before it is detected by an avalanche photodiode (APD) with 10 MHz bandwidth. In this way, the power of the lock light stays below the noise level of the APD. By changing the EOM's driving frequency we are able to control the cavity detuning relative to the atomic resonance.\par
Excitation to the $30D_{5/2}$ Rydberg state is done via two-photon absorption from the $5S_{1/2}$ ground state via the intermediate $5P_{3/2}$ state, Fig.~\ref{fig1}c), using two counter-propagating beams transverse to the cavity axis. For driving the lower transition (pump beam), we use light from the same laser as the probe beam, but controlled with a different AOM, allowing the pump beam to have a detuning relative to the probe. The transition from the intermediate to the Rydberg state (the control beam) is driven with light generated by frequency-doubling a 960 nm wavelength amplified external cavity diode laser (ECDL) using an LBO crystal inside a home-made bow-tie cavity. The frequency is controlled and set to a detuning of $\theta_2=-31.8~\mathrm{MHz}$ by Rydberg EIT spectroscopy in a Rb vapor cell \cite{Abel2009}. An AOM controls the power (120 mW maximum) of this light. Both transverse beams are shaped similarly ,with beam waists of $w_z\sim 700~\mathrm{\mu m}$ along the cavity axis and $w_r\sim 125~\mathrm{\mu m}$ in the vertical direction in order to match the density profile of the atoms in the cavity mode.

\subsection{Experimental procedure}
The experiment starts by loading $N\sim10^6$ Rb$^{87}$ atoms into a MOT where they are cooled down to a temperature of $T\sim10$ $\mu K$ by optical molasses. Lasers are then switched off and after a free fall of the cloud of 2 ms, the initial time for our measurements is set, $t=0$. The overlap between the atom cloud and the cavity mode is optimized by maximizing the normal mode splitting $\Delta_\mathrm{nms}$, shown in Fig.~\ref{fig2}, keeping the transversal beams off. Then, using $\Delta_\mathrm{nms}=2g_0\sqrt{N}$, we derive the effective number of atoms in the cavity mode, $N\sim7.300$.\par
\begin{figure}[t]
\includegraphics{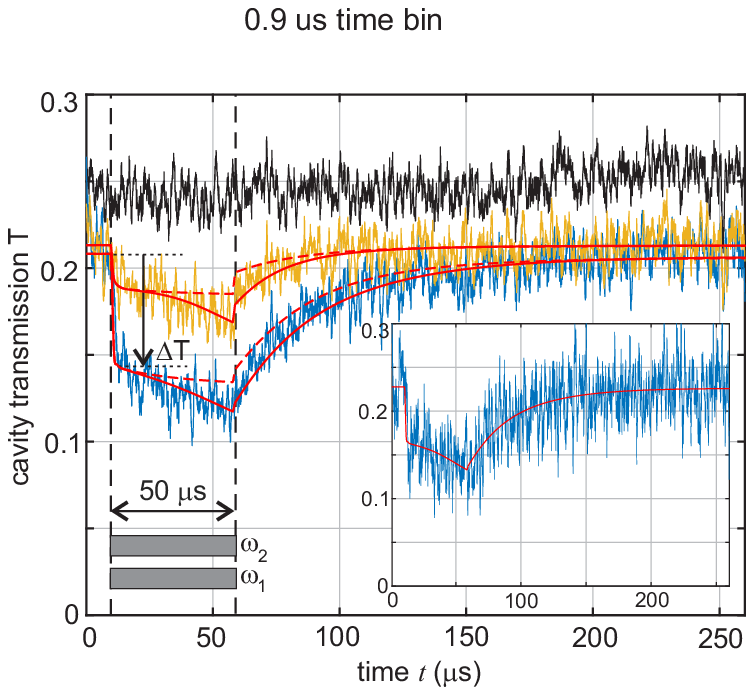}
\caption{\label{fig3} Time signals of the cavity transmission. Atoms are excited to Rydberg states with $\Omega_2=2\pi\times7.6~\mathrm{MHz}$, $\Omega_1=2\pi\times0.9~\mathrm{MHz}$ (yellow) and $\Omega_1=2\pi\times2.9~\mathrm{MHz}$ (blue). The black curve corresponds to $\Omega_1=2\pi\times2.9~\mathrm{MHz}$ and $\Omega_2=0$ and is vertically shifted ($+0.05$) for clarity. The data in the main figure have been averaged over 10 repetitions of the experiment, the inset (same axes as the main figure) shows a single measurement, i.e. it is recorded in real-time. The solid and dashed red lines are fitted simulations of the master equation with and without superradiance, respectively.}
\end{figure}

Similar to \cite{Boddeda2016,Ningyuan2015,Sheng2017}, we measure a cavity Rydberg EIT signal (Fig.~\ref{fig2}) and analyze the data by fitting them to the steady-state equations, from which we determine the Rabi frequency $\Omega_2=2\pi\times5.9~\mathrm{MHz}$ driving the upper transition and the broadened linewidth $\gamma_R=2\pi\times100~\mathrm{kHz}$ of the Rydberg state, compared to the natural linewidth $\Gamma_r=2\pi\times8.3~\mathrm{kHz}$ \cite{arc}. For the measurements shown in Fig.~\ref{fig3}, the parameters were determined in a separate measurement as $\gamma_R=2\pi\times650~\mathrm{kHz}$, $\Omega_2=2\pi\times7.6~\mathrm{MHz}$, and $N\sim6.500$. We attribute the larger linewidth to additional broadening caused by stray electric and magnetic fields. Considering the radius of the atom cloud to be $500~\mu m$, we estimate a maximum atomic density of $\rho=2\times10^9~\mathrm{cm}^{-3}$.\\

For measuring the internal state dynamics of the atoms, we fix the probe detuning at $\theta_p=+24~\mathrm{MHz}$, tagged in Fig.~\ref{fig2}. The detuning of the transverse pump beam is fixed at $\theta_1=+31.8~\mathrm{MHz}$ in order to fulfill the two-photon resonance condition with the coupling beam which has a detuning of $\theta_2=-31.8\mathrm{MHz}$. We turn on the probe beam at $t=0$ and wait for 10 $\mu s$ for the cavity field to equilibrate. Then, we turn on both transverse beams simultaneously for a duration of 50 $\mu s$.\par
Probe light transmitted through the cavity has a power of $\sim$ 2 nW. This corresponds to $\sim$ 200 photons inside the cavity per lifetime $\tau=1/\kappa$ of the intracavity light field. The number of photons is thus small compared to the number of ($N\sim6.500$) atoms, making the weak probe approximation well justified.
\section{Results}
\subsection{Excitation dynamics}
We analyze the transmission of the probe laser through the cavity. The APD signal is electronically low-pass filtered with a timescale of $0.9~\mathrm{\mu s}$. The resulting curve is shown in the inset of Fig.~\ref{fig3}. Before the transverse beams are turned on at $t=10~\mathrm{\mu s}$, the signal corresponds to the steady-state value of the atom-cavity system driven at $\theta_p=+24~\mathrm{MHz}$ (see Fig.~\ref{fig2} for comparison). When the transverse beams are turned on, a fraction ($1-P$ in Eq. \ref{CavTrans}) of the atoms is excited to the Rydberg level. This fraction no longer interacts with the cavity and thereby reduces the normal mode splitting. As a result, the transmission signal drops by $\Delta T$ to a value corresponding to the new atom-cavity steady-state including Rydberg excitation. Rabi oscillations between the ground and Rydberg state are not observed in this experiment due to decoherence. The observation that the signal goes back to its original value after the transverse beams have been switched off shows that the atom loss rate from the cavity is negligible: almost all atoms will eventually reach the ground state by spontaneous decay. We only take loss due to ionization from the Rydberg state with a rate of $\sim 500~\mathrm{Hz}$ \cite{Beterov2007} into account.\par
In order to analyze the excitation process and subsequent decay quantitatively with more precision, we average the recorded traces over 10 repetitions of the experiment. Fig.~\ref{fig3} shows averaged traces for different values of $\Omega_1$ and a fixed value of $\Omega_2$. Increasing $\Omega_1$ leads to a larger value of the Rabi frequency $\Omega_\mathrm{eff}$ and a smaller detuning $\vartheta_\mathrm{eff}$ of the effective two-level system, and thus to
more atoms excited to the Rydberg level, respectively a further reduced cavity transmission, i.e. a larger $\Delta T$. To discern the internal state dynamics from motion out of the cavity mode due to photon recoil, we also measure the transmission when the coupling laser is off (i.e. $\Omega_2=0$), the pump laser Rabi frequency is $\Omega_1=2\pi\times2.9~\mathrm{MHz}$, and the detuning is $\theta_1=+31.8~\mathrm{MHz}$. The resulting transmission (black curve in Fig.~\ref{fig3}) is flat, showing that light-scattering of the pump laser does not contribute to the observed signal. Moreover, recoil from two-photon absorption can be excluded, as few two-photon scattering processes occur on the observed time-scale due to the long lifetime of the Rydberg states compared to the $5P_{3/2}$ state. The yellow curve in Fig.~\ref{fig3} and the simulations reveal also a second, rapid jump, when the transverse beams are switched off at $t=60~\mathrm{\mu s}$. This small jump is due to the fact that the coupling laser generates a detuned cavity Rydberg EIT resonance, as described by Eq. (\ref{Susc3}). For the experimental parameters $\theta_\mathrm{eit}=-7.8~\mathrm{MHz}$ and $\theta_p=24~\mathrm{MHz}$ the occurrence of this EIT resonance modifies the atom-cavity's steady state, as long as the coupling beam is switched on. Correspondingly, a jump also occurs when the laser is switched on at $t=10~\mathrm{\mu s}$, and partially contributes to $\Delta T$. This is taken into account in the simulations.
\subsection{BBR-induced transitions}
We further want to analyze the time signal: the blue and yellow curve in Fig.~\ref{fig3} corresponds to an effective Rabi frequency of $\Omega_\mathrm{eff}=2\pi\times350~\mathrm{kHz}$ and $\Omega_\mathrm{eff}=2\pi\times109~\mathrm{kHz}$, respectively. The steady-state of the effective two-level system is reached on a timescale given by the broadened linewidth $\gamma_R=2\pi\times650~\mathrm{kHz}$, in our case $\sim2~\mathrm{\mu s}$, matching the observed timescale of the steep decrease $\Delta T$. However, it is evident that the system has not yet reached its steady-state after the decrease, as the signal is continuously decreasing during the 50 $\mu s$ pulse. This is explained by shelving of the atoms into other long-lived Rydberg states populated from the $30D_{5/2}$ state due to BBR. Similarly, $42\%$ of the spontaneous decay from the $30D_{5/2}$ state leads to other states than $5P_{3/2}$; however, the contribution of these to shelving is negligible due to their short lifetimes. Nevertheless, our simulations include most decay products from the $30D_{5/2}$ state (at least $0.5\%$ contribution to the corresponding total decay rate) until they reach the ground state. BBR induced decay products are included only for the $30D_{5/2}$ state and its closest neighbours. In total, we simulate the dynamics of 347 states following Eq. (\ref{RateEqs}). Even though the atom number and the cavity detuning suffer shot-to-shot fluctuations within $N=6.500\pm 200$ and $\theta_\mathrm{cav}=-1.3\pm 0.5~\mathrm{MHz}$, in our simulations we fix $N$ and only optimize $\theta_\mathrm{cav}$ within its uncertainty. Therefore, the red dashed lines in Fig.~\ref{fig3} are, apart from $\theta_\mathrm{cav}$, without free parameters. They quantitatively agree with the experimental curves, but do not explain the observed slow decrease of transmission during the pulse. Apparently, more atoms are taken out of the $30D_{5/2}$ state than what BBR-induced transitions and spontaneous decay can explain.
\subsection{Superradiance and dipole-dipole interactions}
The discrepancy can be ascribed to superradiance, which we consider here to occur in the transitions from the $30D_{5/2}$ state to the $31P_{3/2}$ state and the $28F_{7/2}$ state. SR is principally possible in these transitions as the longest lengthscale of the system (i.e. the radius of the cloud  with $R=0.5~\mathrm{mm}$) is smaller than the corresponding transition wavelengths of $3.6~\mathrm{mm}$ (to $31P_{3/2}$) and $1.5~\mathrm{mm}$ (to $28F_{7/2}$).\par
\begin{figure}[t]
\includegraphics{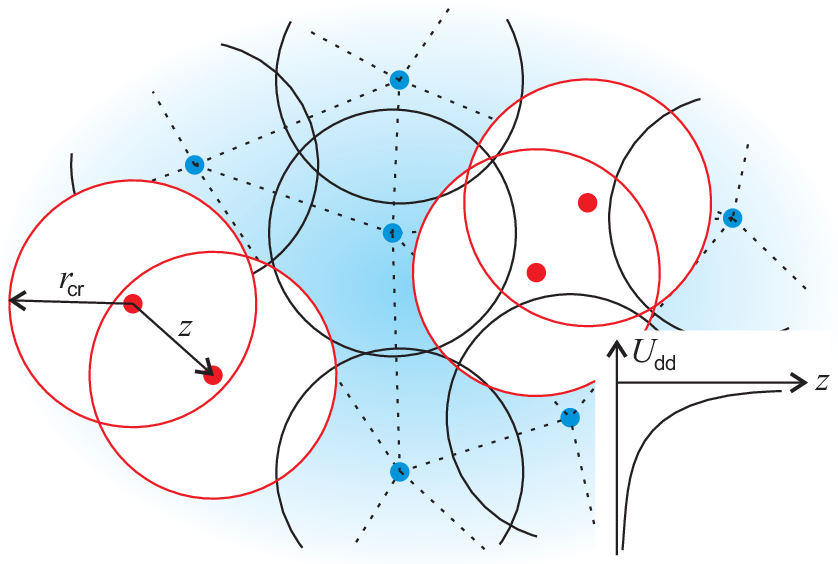}
\caption{\label{fig4} Atoms (in red) with distance $z$ to their next neighbour smaller than the critical radius $r_\mathrm{cr}$ are tuned out of resonance due to the dipole-dipole interaction potential $U_\mathrm{dd}(z)$. They are not part of the extended coherent many-body state (Dicke state)  which is formed during superradiance, as indicated by the dashed lines connecting the atoms (in blue).}
\end{figure}
\begin{figure}[t]
\includegraphics{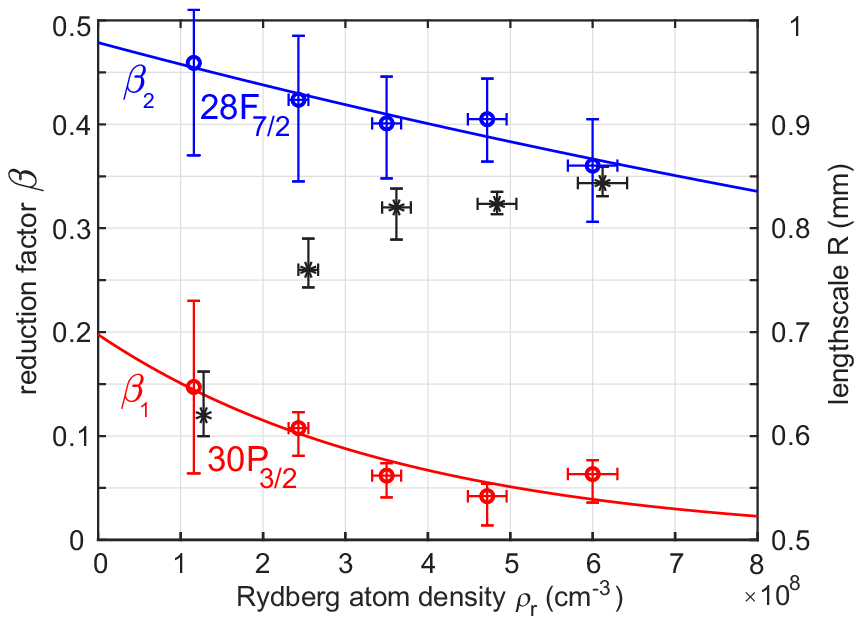}
\caption{\label{fig5} Circles (left axis) show the reduction factors $\beta_{1}$ and $\beta_{2}$ obtained from fitted time signals (Fig.~\ref{fig3}), corresponding to the superradiant transitions from $30D_{5/2}$ to $31P_{3/2}$ and $28F_{7/2}$. Vertical error bars indicate the $95\%$ confidence interval. Horizontal error bars correspond to a $5\%$ uncertainty in the atom number. The solid lines are exponential fits to these data points, following Eq. (\ref{beta_fit}) for varying Rydberg atom density $\rho_r$. The fits deliver critical radii of $r_\mathrm{cr,1}=8.6~\mu m$ and $r_\mathrm{cr,2}=4.7~\mu m$. Stars (right axis) show the lengthscale $R$ of the superradiant cloud, also derived from the fitted time signals (Fig.~\ref{fig3}). The decrease of $R$ for small density is caused by the nonlinear dependence of $\rho_r$ on $\Omega_1$ and the Gaussian profile of the exciting laser beams. Data for $R$ are shifted to the right by one data width for clarity.}
\end{figure}

On the other hand, Dicke states, responsible for SR, are extended coherent many-body states. This makes them susceptible to dephasing due to inhomogeneities in the transition energies across the sample \cite{Vasily2005}. In turn, Rydberg atoms can bring up more dephasing due to long-range dipole-dipole interactions between pairs of Rydberg atoms in different states \cite{Zhou2016}. Unlike for inhomogeneous electric or magnetic fields, such interaction-based dephasing strongly depends on the distance between neighbouring atoms; thus, on the Rydberg atom density $\rho_r$. We model the impact of dipole-dipole interactions on SR by assuming that atoms with a neighbouring atom closer than some critical radius $r_\mathrm{cr}$ are tuned out of resonance and thus, do not contribute to SR, as indicated in Fig.~\ref{fig4}. This critical radius is defined by the distance where the dipole-dipole interaction potential 
\begin{equation}\label{dd_int}
U_\mathrm{dd}=C_3/z^3,
\end{equation}
with interatomic distance $z$ and $C_3$ coefficient, equals the broadened linewidth of the $30D_{5/2}$ state. The probability $p(\rho_r,r_\mathrm{cr})$ that the nearest neighbour of an arbitrarily chosen Rydberg atom is within a spherical volume with radius $r_\mathrm{cr}$ is derived from the nearest neighbour distribution function as 
\begin{equation}\label{nextneighbour}
    p(\rho_r,r_\mathrm{cr})=1-\beta_{rl}=1-e^{-\frac{4\pi}{3}\rho_r  r_\mathrm{cr}^3}.
\end{equation}
We reduce the number of superradiant atoms in our simulation by replacing $N_r$ in Eq. (\ref{SupDiff}) with  $\beta_{rl} N_r$. The reduction factor $\beta_{rl}$ depends on the levels $r$ and $l$ of the superradiant transition via the corresponding $C_3$ coefficient. In our simulation we introduce and fit two values $\beta_{1,2}$ for the two transitions from $30D_{5/2}$ to $31P_{3/2}$ and $28F_{7/2}$, respectively. We furthermore fit the lengthscale $R$ of the superradiant atom cloud via the cooperativity factor $C_{rl}$. We perform the fit for 5 different values of $\Omega_1$ and show the results for $\beta_1$, $\beta_2$ and $R$ in Fig.~\ref{fig5} as function of Rydberg atom density. Here, we define $\rho_r=P_r\rho$ where $P_r$ is the steady-state probability for an atom to be in the main Rydberg state calculated by simulating a closed 3-level system with our experimental parameters, and $\rho$ is the peak density in the center of the cloud. Correspondingly, $P_r$ is calculated for the maximum light intensity in the center of the exciting beams.
The obtained values for the reduction factors $\beta_{1,2}$ are fitted with the function
\begin{equation}\label{beta_fit}
    \beta_{1,2}=A_{1,2}e^{-\frac{4\pi}{3}\rho_r  r_\mathrm{cr,1,2}^3},
\end{equation}
following the theoretical expectation (\ref{nextneighbour}), and including an additional reduction amplitude $A_{1,2}$, see solid lines in Fig.~\ref{fig5}. The factors $\beta_{1,2}$ are proportional to the probability that an arbitrary Rydberg atom is ready to participate in the corresponding superradiant decay. We fit both parameters $A_{1,2}$ and $r_\mathrm{cr,1,2}$. For the critical radii, we obtain values of $r_\mathrm{cr,1}=8.6~\mu m$ for the transition to $31P_{3/2}$ and $r_\mathrm{cr,2}=4.7~\mu m$ for the transition to $28F_{7/2}$. Their values can now be compared with those deduced theoretically from the dipole-dipole interaction potential under the condition $U_\mathrm{dd}=\gamma_r/(2\pi)=650~\mathrm{kHz}$. We use $C_3=-0.4858~\mathrm{GHz~\mu m^3}$ for the potential between an atom in the $30D_{5/2},m=5/2$ state and an atom in the $31P_{3/2},m=3/2$ state, and $C_3=-0.1654~\mathrm{GHz~\mu m^3}$ for the potential between an atom in the $30D_{5/2},m=5/2$ state and an atom in the $28F_{7/2},m=7/2$ state \cite{Weber2017}. Thus, the theoretical critical radii are $r_\mathrm{cr,1}^\mathrm{th}=9.1~\mu m$ and $r_\mathrm{cr,2}^\mathrm{th}=6.3~\mu m$. The experimentally determined values are slightly smaller. This might be caused by the fact, that the atoms are spread over the magnetic sublevels of the $30D_{5/2}$ state. This opens the possibility of pair-interactions between atoms with different $m_J$ combinations whereas the theoretical value corresponds to only one combination. Fitting the amplitude $A_{1,2}$ of the exponential functions takes into account that the  number $N_r$ of superradiant Rydberg atoms in Eq. (\ref{SupDiff}) can be further reduced, for instance by the spread of occupation over the magnetic sublevels of the $30D_{5/2}$ state and their decay to different magnetic sublevels of the final Rydberg states.
\par
The fits of the time signals (Fig.~\ref{fig3}) deliver also the lengthscale $R$ of the superradiant cloud. We observe in Fig.~\ref{fig5} that $R$ decreases as the Rydberg atoms get more dilute. This is caused by the nonlinear dependence of the excitation probability $P_r$ on the Rabi frequency, which scales like $P_r\propto\Omega_1^2$ for weak driving, according to the effective 2-level system. Thus, excitation is suppressed at the outer regions of the Gaussian laser beams where the Rabi frequency is smaller than in the center. The resulting Rydberg atom cloud is smaller than the Gaussian profile of the Rabi frequency. If, however, the peak Rabi frequency at the beam center is larger, the excitation probability scales linearly as $P_r\propto\Omega_1$. In this case, the excited Rydberg atom cloud matches the Gaussian profile of the Rabi frequency. For a Gaussian intensity profile and in the limit of $\Omega_1\rightarrow 0$, corresponding to the limit of $\rho_r\rightarrow 0$, the lengthscale of the excited Rydberg atom cloud gets reduced by a factor of $1/\sqrt{2}$, matching the observation in Fig.~\ref{fig5}.\par

\section{Conclusions}
We have shown that, taking advantage of the atom number dependence of the collective atom-cavity coupling, it is possible to observe Rydberg dynamics during excitation and decay. With $\sim$ 6.500 atoms in the cavity, we observe Rydberg dynamics by monitoring the light transmitted through the cavity while weakly driving it at the side of fringe of the normal mode splitting. Two laser fields propagating transversally to the cavity axis pump the atoms to the $30D_{5/2}$ Rydberg state via a 2-photon transition. At first, the cavity transmission rapidly decays, corresponding to the transfer of a fraction of atoms to the Rydberg state. The rapid decrease is followed by a slower decay which is described by BBR-induced population redistribution of atoms into other Rydberg states, where they are "shelved" before they decay back to the ground state. We simulate the atomic dynamics with rate equations taking several decay channels including BBR-induced transitions into account. The simulation fully agrees with the data only when superradiance is included. Most importantly, it is observed that superradiance is suppressed for growing Rydberg atom density. This effect is attributed to an inhomogeneous dephasing of the many-body Dicke state caused by dipole-dipole interactions between atoms in neighbouring Rydberg states. A model is introduced that effectively reduces the atom number available for superradiance, depending on the next neighbour separations compared to the interaction range of the dipole-dipole interactions.\par 

It is noteworthy that the observation of superradiance in the decay to neighbouring Rydberg states requires, on one hand, sufficiently large numbers of Rydberg atoms for enhancement. This is the reason why superradiance is elusive in experiments with few atoms. On the other hand, it requires low enough densities to avoid dipole-dipole interactions. Our experiment with a dilute MOT features Rydberg atom densities corresponding to most probable next neighbour separations of $d=(2\pi\rho_r)^{-1/3}=6~\mathrm{\mu m}~...~12~\mathrm{\mu m}$, which is just in the right range to probe the growing influence of long range dipole-dipole interactions. Thus, our observations contribute to resolve a recent discussion in the Rydberg community concerning the occurrence of superradiance and the role of dipole-dipole interactions in this context.\par

Our observations are real-time in the sense that each experimental run delivers a curve showing the complete dynamics. In principle, no sampling or averaging over several runs is required. Thus, the method can potentially be used for generating feedback to the system by, for instance, controlling the length of the exciting pulse depending on the current cavity transmission. This perspective is particularly interesting for cavities with higher finesse featuring atom resolution and spin-squeezing. Also, by improving the ratio between the effective Rabi frequency and the rate of decoherence it will be possible to resolve Rabi oscillations in real-time. 

\begin{acknowledgments}
We acknowledge helpful discussions with Manuel Kaiser and technical support by Dalila Rivero Jerez. The project was funded by the Deutsche Forschungsgemeinschaft (DFG, German Research Foundation) - 422447846 and by ColOpt—EUH2020 ITN 721465.
\\
E.S. conducted the experiment and collected the data. E.S. and S.S. analyzed the data, developed the physical model and wrote the paper, equally. S.S. conceptualized the work. Ph.W., P.W. and E.S. constructed the experimental setup including the reference cavity, the locking scheme for the science cavity, and the laser frequency lock using Rydberg EIT spectroscopy. 
\end{acknowledgments}

\bibliography{Ry_dynamics}

\end{document}